\documentclass[journal]{IEEEtran}

\usepackage{graphicx}
\usepackage{cite}
\usepackage{booktabs}
\usepackage{multirow}
\usepackage{url}
\usepackage{amsmath,amssymb} 
\usepackage[hidelinks]{hyperref}

\begin{document}

\title{Entanglement Swapping Scheduling for Quantum Repeater Chains under Decoherence during Classical Communications}

\author{
Jorge Benito Díaz,
Andrés Agustí,
Javier Faba,
Laura Ortiz,
Luis Robledo,
Vicente Martín
\thanks{Jorge Benito Díaz and Andrés Agustí are with Universidad Politécnica de Madrid.}
\thanks{Javier Faba and Laura Ortiz are with DATSI at Universidad Politécnica de Madrid}
\thanks{Vicente Martín is with DLSIIS at Universidad Politécnica de Madrid}
\thanks{Luis Robledo is with DFT at Universidad Autónoma de Madrid}
}

\maketitle

\begin{abstract}
First-generation quantum repeaters are  essential in distributing entanglement over long distances since  they could mitigate losses in optical fiber. Repeaters split the total distance into shorter, easier to generate links. Then, entanglement swapping establishes long range entanglement by performing local measurements at the repeaters and heralding their results. 
In this work, we develop an analytical model that estimates the impact of decoherence during heralding for a given protocol.

Furthermore, we identify a competition between two phenomena: as repeater count increases, so does entanglement generation, but the additive impact of memory decoherence on final entanglement rate increases too. Therefore, we find that an optimal number of repeaters exists. These findings could lead the design of future quantum networks, both in the inter-repeater distance and swapping timings.
\end{abstract}


{\footnotesize
This work has been submitted to the IEEE for possible publication. Copyright may be transferred without notice, after which this version may no longer be accessible.
}
\section{Introduction}
\label{sec: introduction}


First-generation quantum repeaters serve as essential devices for quantum communications. They enable entanglement distribution across network nodes overcoming the exponential attenuation in optical fiber and unlocking quantum capabilities that surpass classical networking paradigms. These include quantum key distribution (QKD), blind quantum computing, distributed quantum computing, quantum-enhanced sensing or clock synchronization \cite{BennettBrassard1984,BroadbentFitzsimonsKashefi2009, CiracEkertHuelgaMacchiavello1999,GiovannettiLloydMaccone2004,JozsaAbramsDowlingWilliams2000}. Quantum repeaters rely on two fundamental operations: entanglement generation (EG), which creates short-range entanglement between physically connected adjacent nodes, and entanglement swapping (ES), which uses adjacent entangled links at intermediate nodes to extend the entanglement distance across previously unlinked segments.

By enabling long-distance entanglement distribution, quantum repeaters pave the way for the quantum internet, where chains of repeaters --- deployed either via terrestrial fiber networks or free-space satellite links --- will interconnect geographically separated metropolitan area networks. In such quantum networks, optimizing entanglement distribution along the repeater chain becomes a crucial challenge. We identify three main components to said challenge. First, a model that summarizes the physical state of the chain must be introduced, and it must correctly compute the evolution of the chain under any EG or ES operations. Second, a policy must govern under which conditions these EG and ES operations are executed along the chain, that is, a policy must drive the chain dynamics. Third, we must identify a performance metric that correctly summarizes our interests in the output of those dynamics.

Regarding performance metrics, previous literature proposes metrics that can be broadly sorted into three groups \cite{mai2026swappingcapacityquantumrepeater,brand2020,shchukin2019}. Some of them, such as the expected time to complete the protocol, quantify the number of long-range end-to-end (E2E) entangled pairs generated. Other metrics, such as the final state fidelity to a Bell pair, quantify the quality of a pair. However, there is third metric that summarizes both phenomena. Logarithmic negativity is an additive entanglement measure over multiple entangled pairs. Therefore, protocols with high rate and low quality can be compared against low-rate and high-quality protocols by adding this measure over all the entangled pairs generated by both protocols over a common period of time and dividing by that time. This is the entanglement rate that we use in this work.

Regarding the network model, there are some phenomena that are central to its dynamics. As mentioned above, EG and ES are its main operations. However, we must point out that both of those operations are probabilistic and, under successful completion, require the heralding of said success to other constituents in the network. This messaging introduces a delay that not only reduces the entanglement rate by slowing the protocol down, but reduces the entanglement quality because of the decoherence on the memories. To rigorously model the impact of classical signaling latency, we introduce the concept of exposure, which quantifies the temporal decoherence suffered by a number of entangled pairs while waiting for classical confirmation signals, which is one of the main contributions of this work.

Finally, regarding the policies with which to sort EG and ES operations, they can be categorized into two main types: non-reactive and reactive. Non-reactive policies adhere to a pre-established sequence of operations along the chain, whereas reactive policies dynamically execute operations based on the real-time network state. Although reactive policies have more information available to carry out their operations, they must wait for classical signaling to propagate through the network. This introduces additional delays that slow the protocol and degrade entanglement. It is not clear if their awareness of the network state provides an advantage that overcomes their delays. Moreover, the cost of classical signal propagation during the late stages of ES rounds is difficult to compute. Therefore, we choose to study policies that are reactive only in EG operations, where the cost is low and easy to compute, while they are non-reactive in the ES operations.

With these tools, we provide an analytical and hardware-agnostic framework that allows us to extract novel conclusions that may guide the design of a future Quantum Internet architecture in the presence of decoherence in quantum memories while accounting for the delays in heralding. In addition to this model, our main findings include the characterization of the optimal number of quantum repeaters for a given chain and the order of their ES operations, as well as determining the range of values for coherence times and inter-repeater distance over which entanglement distribution is possible.

This work is structured as follows: Section~\ref{sec: theoretical_model} introduces the theoretical model, including the physical model and operational primitives, the ES scheduling strategies considered, and the temporal and performance framework. Section~\ref{sec: results} presents and discusses the results obtained and derives their main architectural implications. Finally, Section~\ref{sec: conclussion} summarizes the main conclusions and outlines future research directions.

\section{Theoretical Model}
\label{sec: theoretical_model}

\subsection{Physical Model and Operational Primitives}
\label{subsec: physical_model}

We consider a homogeneous linear chain consisting of $N$ quantum repeaters, 
connecting two end nodes, 
over a total distance $L_{\mathrm{tot}}$ and divided into $N+1$ elementary links of equal length $L$. Each elementary link supports a single quantum channel. The probability $p(L)$ of successfully generating an entangled pair in one attempt is equal to  the baseline EG probability modulated by an exponential factor accounting for photon losses during transmission through the optical fiber. 

When an EG attempt succeeds, the resulting pair is stored in quantum memories\footnote{Each intermediate repeater contains one memory qubit
associated with each adjacent elementary link, resulting in a total
of two memory qubits.} located at the nodes connected by the corresponding elementary link. The quantum memories are characterized by an effective decoherence time
$\tau$, related to the coherence times of each memory. Each successfully generated entangled pair is characterized as a Werner state with initial parameter $w_0$.

Since EG is heralded, a pair cannot be considered available for subsequent operations until the corresponding classical information has reached the relevant nodes. The pair is stored in memories immediately after generation and therefore already undergoes decoherence during this heralding delay. Additionally, once the successful generation has been confirmed, the pair may remain stored for an additional time while the other elementary-links continue to be generated. Hence, the total memory degradation accounts for both the classical-communication delay associated with the heralded EG attempt and any subsequent waiting time before the pair is used. We model this degradation as an exponential time-decay of the Werner parameter.

Once a repeater holds two entangled pairs, 
one per memory,
it can perform an ES operation, which succeeds with probability $q$. After success, swapping two Werner states multiplies their parameters to create a longer range state. We refer to a single ES operation performed at one repeater as a $1$-ES operation. More generally, an $n$-ES operation consists of $n$ $1$-ES operations performed simultaneously at adjacent repeaters.
Assuming independent ES outcomes and assuming that the probability of $1$-ES of each repeater is $q$, the probability that the $n$-ES operation succeeds is $q^n$.

As in heralded EG, completing an $n$-ES operation requires local processing followed by classical communication of the measurement outcomes from the repeater nodes to the involved nodes. Consequently, ES operations introduce non-negligible time delays, and the entangled states stored in quantum memories continue to decohere throughout the associated waiting time. The specific communication distances and accumulated storage times depend on how the $n$-ES operations are organized within the composition phase and must therefore be determined separately for each scheduling strategy. 

\subsection{Entanglement-Swapping Scheduling Strategies}
\label{subsec: scheduling_strategies}

Within
policies introduced in Section~\ref{sec: introduction}, we restrict our analysis to prepare-and-swap strategies, in which elementary-link generation and entanglement swapping are performed in two separate phases: each elementary link attempts EG until its first successful pair is obtained, and the composition phase begins only after all elementary links are ready. Therefore, no ES operation is attempted during the generation phase. This assumption excludes purely reactive strategies, such as swap-as-soon-as-possible (swap-ASAP), from the scope of the present work. However, it enables an analytical characterization of the E2E entanglement distribution probability, and the timing and memory-degradation contributions associated with each scheduling strategy.

Under this prepare-and-swap framework, we consider the following three scheduling strategies:

\begin{itemize}

 \item \textbf{Binary-Tree scheduling} \cite{Briegel1998,ghaderibaneh2022}: the chain is composed hierarchically through successive stages of $1$-ES operations. In the first stage, these operations are performed in parallel at selected nonadjacent repeaters to connect neighboring elementary links. In each subsequent stage, the entangled segments produced earlier are composed likewise until the two end nodes are connected. In this work, the tree is not required to be balanced; therefore, some entangled segments may remain stored across one or more stages until a compatible neighboring segment becomes available for swapping.
 
    \item \textbf{Parallel scheduling} \cite{abane2025}: all ($N$) repeaters perform their ES operations simultaneously, so that the complete chain is composed through a single $N$-ES operation. This strategy maximizes parallelism and completes the swapping process in a single composition stage.
    
    \item \textbf{Hybrid scheduling}: the chain is first divided into $B$ blocks of adjacent elementary links. Within each block, the corresponding repeaters perform an $n$-ES operation where $n$ is determined by the number of repeaters contained in each block. In a second stage, the resulting segments are composed in a final single $n$-ES operation. This strategy combines the simultaneous composition of the parallel strategy with the hierarchical structure of the binary-tree approach.
\end{itemize}

Although all three strategies ultimately connect the same set of elementary links, they differ in the number and organization of composition stages, and in the communication distances required to complete them. These differences result in strategy-dependent execution times and memory-exposure patterns, which are characterized in the following subsection.

\subsection{Temporal Model, Memory Decoherence and Performance Metric}
\label{subsec: times_decoherence_metric}

We now introduce the temporal model used to quantify the total memory exposure induced by the EG and ES phases. Since Werner parameters combine multiplicatively under ES, while memory decoherence acts exponentially on each stored state, the relevant temporal quantity is not a single elapsed time but the cumulative sum of storage times experienced by the entangled states contributing to the final E2E pair. We refer to this quantity as the memory exposure $\Theta$. This quantity has two contributions: $\Theta_{\mathrm{gen}}$ and $\Theta_{\mathrm{swap}}$ --- which is strategy-dependent --- accounting for EG and ES phases, respectively. 

Since each elementary link attempts EG independently until its first success, the number of generation rounds required by each link follows a geometric distribution with success probability per trial $p(L)$, mentioned in Subsection~\ref{subsec: physical_model}. We neglect local processing delays and assume that the duration of one round of each generation attempt is determined by the classical propagation delay over the elementary link. The generation phase ends when the last elementary link achieves its first successful EG, so its duration is determined by the slowest generation realization in that execution. After the last elementary link becomes ready, this information must propagate across the chain before the ES phase begins. We account for this additional delay using a conservative configuration where the last EG success occurs at one end of the repeater chain and its information must propagate across the whole chain. The generation-phase memory exposure $\Theta_{\mathrm{gen}}$ therefore accounts for both the asynchronous waiting time accumulated by pairs generated before chain preparation is complete and the global readiness delay experienced by all stored pairs. Its expectation is obtained by averaging these storage times over the stochastic elementary-link generation histories.

During the swapping phase, several ES operations may belong to the same composition stage, and their heralding information is assumed to propagate in parallel. The duration of this stage is therefore determined by the longest distance required to complete it, called as its communication span. The memory exposure associated with a stage is consequently given by the corresponding communication time multiplied by the number of pairs that remain stored while the information propagates. Finally, the total swapping-phase memory exposure $\Theta_{\mathrm{swap}}$ is obtained by summing the exposures over all composition stages. The communication span and storage pattern of each stage are determined by the scheduling rules introduced in Section~\ref{subsec: scheduling_strategies} and therefore differ among the three strategies.

For a successful execution, the Werner parameter of the final E2E state is obtained by multiplying the initial contribution $w_0^{N+1}$ by an exponential decay determined by the total memory exposure $\Theta$ relative to the memory decoherence time $\tau$. For a given generation history, we use the logarithmic negativity to quantify the entanglement of the E2E Werner state conditioned on successful completion of the swapping phase. Since all strategies require $N$ independent ES operations, this contribution is weighted by the common success probability $q^N$. The expected distributed entanglement $\mathcal{E}$ is then obtained by averaging the resulting logarithmic negativity over all possible generation histories, accounting for the probability of each history and the round in which chain preparation is completed. 

While the expected distributed entanglement quantifies the average entanglement delivered per protocol execution, a complete performance metric requires evaluating the rate of entanglement distribution over time. To this end, we use the expected distributed entanglement per unit time, named as the entanglement rate $\mathcal{R}$. This quantity is defined as the ratio of the expected distributed entanglement to the expected execution time of the protocol. This expected execution time comprises the average duration of the generation phase and each subsequent swapping stage, weighted by the probability that all preceding swapping stages have succeed. For instance, in a multi-stage strategy, if an ES failure occurs at an intermediate composition stage, the swapping time accounted for that outcome includes only the time elapsed up to the point of failure and that failed execution contributes zero distributed entanglement. Therefore, all subsequent swapping stages are skipped, reducing the time consumed by that execution. We refer to this as an early-abort policy. In addition, determining when to restart the protocol requires classical communication to broadcast the outcome (success or failure) across the network. We explicitly incorporate this classical signaling overhead --- analogously to the global readiness delay introduced for the EG phase --- by accounting for the maximum transmission time required for all nodes in the chain to receive the status update before a new entanglement distribution attempt can be initiated.

\section{Results and Discussion}
\label{sec: results}

Having constructed the theoretical model, this section examines its quantitative behavior under a representative set of parameter values. Unless otherwise stated, we set the parameters $p_0$, $q$ and $w_0$ to 0.5, 0.8 and 1 respectively. These values are not intended as a direct calibration of a specific experimental platform but rather as an effective reference parameterization for studying the structural behavior of the model. The parameter $p_0$ represents the EG probability in the absence of propagation losses and incorporates local efficiencies associated with state preparation, coupling, detection and heralding. Its value is interpreted as a high but non-deterministic reference, consistent with the probabilistic nature of photonic EG primitives, such as conventional linear-optical Bell-state measurements 
\cite{Azuma2023RMP}. Similarly, $q=0.8$ represents an imperfect but relatively high elementary swapping success probability. Experimental demonstrations of ES and teleportation between non-neighbouring nodes illustrate the physical relevance of this operational regime \cite{Hermans2022TeleportationNetwork}. 
Finally, setting $w_0=1$ allows us to isolate the degradation induced by memory decoherence from the temporal organization of the different scheduling strategies.

To study the spatial dependence of the model, we consider two different scenarios: a metropolitan scenario with $L_{\mathrm{tot}}=60\,\mathrm{km}$ and an intermetropolitan scenario with $L_{\mathrm{tot}}=500\,\mathrm{km}$. The former represents an extended metropolitan scale where direct transmission can remain competitive, whereas the latter represents long-distance intercity connections, where fiber attenuation becomes substantially more severe. For a fixed $L_{\mathrm{tot}}$, varying the number of repeaters $N$ simultaneously modifies the number of elementary links and their length $L$, giving rise to two competing effects. Increasing $N$ shortens the elementary-link distance, thereby increasing the EG probability per round and reducing the expected generation time of each individual link. However, it also increases the number of elementary links that must be prepared and the number of ES operations required to connect the end nodes, thereby reducing the overall swapping success probability and potentially increasing the global preparation time, the accumulated memory exposure, and the expected execution time. 

All results are evaluated using the rate-based performance metric introduced in Section~\ref{subsec: times_decoherence_metric}, including early-abort policy upon detection of an ES failure. For the hybrid strategy, the block parameter $B$ is optimized independently at each point of the parameter space, thus the reported performance corresponds to the best admissible hybrid configuration. We restrict this optimization to non-trivial configurations, since the limiting cases $B=1$ and $B=N+1$ collapse to parallel scheduling. Moreover, additional structural degeneracies occur for short chains: for $N=1$ all strategies trivially reduce to the same single $1$-ES operation, whereas for $N=2$ and $N=3$ some hybrid configurations coincide with binary-tree scheduling.

\subsection{Metropolitan Scenario}
\label{subsec: metropolitan}

\begin{figure}[t]
    \centering
    \includegraphics[width=\columnwidth]{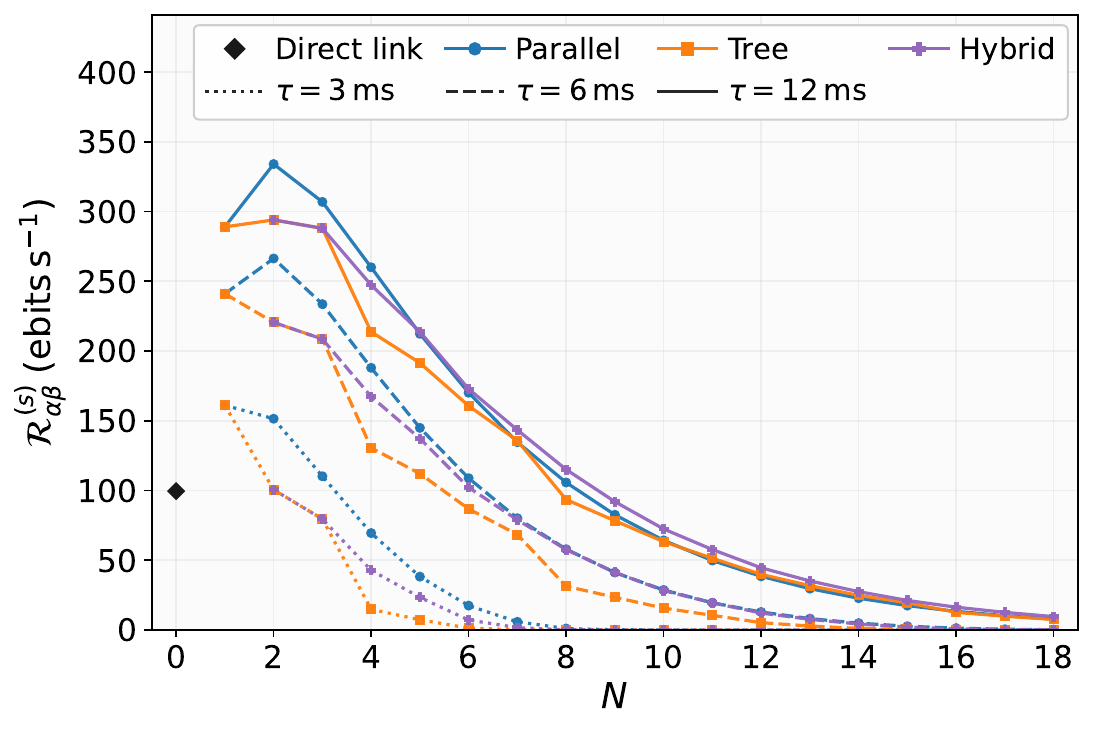}
    \caption{Entanglement rate as a function of the number of intermediate repeaters for the metropolitan scenario, with $L_{\mathrm{tot}}=60\,\mathrm{km}$, $p_0=0.5$, $q=0.8$, and $w_0=1$. Results are shown for three representative memory decoherence times $\tau\in\{3,6,12\}\,\mathrm{ms}$, using the early-abort policy. The direct-link configuration $N=0$ is included as a baseline only for $\tau=6\,\mathrm{ms}$.}
    \label{fig:metro_rate}
\end{figure}

Figure~\ref{fig:metro_rate} shows that the optimal repeater count remains small throughout the memory regimes considered. For $\tau=3\,\mathrm{ms}$, the maximum rate is obtained with a single repeater, while the optimum shifts to $N=2$ for the rest. In all cases, parallel scheduling provides the highest global rate around the optimum.

The effect of memory decoherence becomes increasingly apparent as the number of repeaters increases. For short-lived memories, the rate decreases rapidly beyond the optimum, as the reduction in elementary-link length is unable to compensate for the increasing number of stored states and ES operations. Increasing $\tau$ mitigates this penalty and extends the range of repeater counts in which appreciable rates can be maintained. Nevertheless, even for the maximum $\tau$ considered, the benefit of introducing additional repeaters eventually saturates and the rate decreases with $N$, illustrating that denser repeater deployment is not necessarily advantageous under finite-memory conditions.

The relative performance of the scheduling strategies also changes away from the optimum. While parallel scheduling benefits from completing the composition phase in a single stage and dominates the high-rate region, the hybrid strategy becomes increasingly competitive for larger $N$ and can slightly outperform the parallel strategy in parts of the long-memory regime. However, these crossings occur only after the absolute rate has already fallen below its maximum. The metropolitan results therefore favor short repeater chains, with parallel scheduling providing the best performance in the relevant region around the optimum.

\subsection{Intermetropolitan Scenario}
\label{subsec: intermetropolitan}

\begin{figure}[t]
    \centering
    \includegraphics[width=\columnwidth]{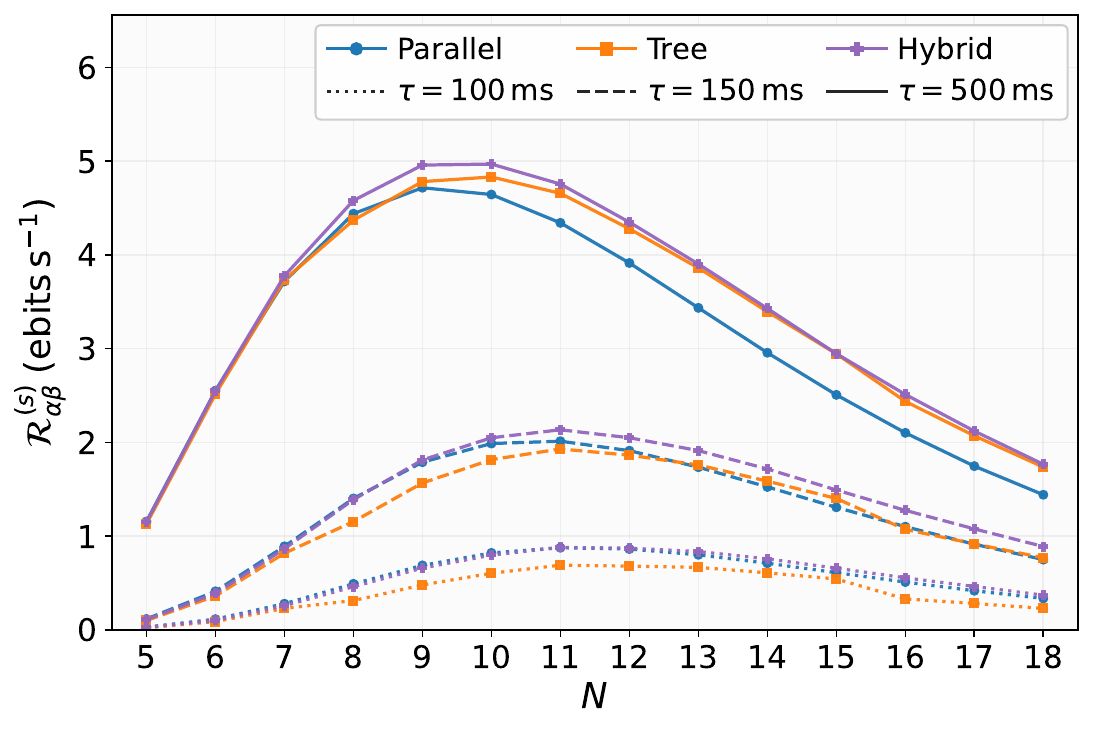}
    \caption{Entanglement rate as a function of the number of intermediate repeaters for the intermetropolitan scenario, with $L_{\mathrm{tot}}=500\,\mathrm{km}$, $p_0=0.5$, $q=0.8$, and $w_0=1$. Results are shown for three representative memory decoherence times $\tau\in\{100,150,500\}\,\mathrm{ms}$, using the early-abort policy.}
    \label{fig:intermetro_rate}
\end{figure}

Figure~\ref{fig:intermetro_rate} shows a qualitatively different behavior from the metropolitan scenario. For all memory regimes considered, the entanglement rate exhibits a clear interior maximum as a function of the number of repeaters. Increasing $N$ initially improves the rate by shortening elementary links but, beyond the optimum, the increasing number of stored states and ES operations, together with the associated memory exposure and swapping-failure probability, eventually outweigh this benefit.

The position of the optimum depends on the memory regime. For short values of $\tau$, the maximum occurs around $N=10$--$11$, whereas for $\tau=500\,\mathrm{ms}$ it shifts towards smaller repeater counts. Longer-lived memories reduce the penalty associated with storing entangled elementary links, making further subdivision of the chain progressively less beneficial. This behavior illustrates the existence of a non-trivial optimum repeater count from the interplay between fiber attenuation, heralding-induced delays and memory exposure, and probabilistic swapping.

The relative performance of the strategies also differs from that in the metropolitan case. The hybrid strategy achieves the highest rate over most of the relevant parameter range, particularly around the optimum for the longer memory regimes, while binary-tree scheduling approaches its performance as $\tau$ increases. Parallel scheduling remains competitive for short chains but becomes increasingly penalized as $N$ grows. The intermetropolitan results therefore show that, over sufficiently long distances, hierarchical organization of the swapping phase can provide a performance advantage while preserving the existence of a well-defined optimal repeater count.

\section{Conclusion and Future Work}
\label{sec: conclussion}

We have developed a hardware-agnostic analytical framework to study the quantitative behavior of ES scheduling strategies in first generation quantum repeater chains under finite memory decoherence times and classical signaling delays. By introducing the notion of memory exposure, the model captures the cumulative decoherence induced by both asynchronous EG and the organization of the swapping phase. An entanglement rate metric has been evaluated. It jointly accounts for the expected distributed entanglement delivered per protocol execution and the expected protocol execution time, including early abort upon detection of an ES failure.

Our results suggest that repeater density should be treated as an architectural design variable rather than as a quantity to be maximized. Shortening the elementary links improves EG, but the associated increase in stored resources, probabilistic ES operations, and heralding delays eventually offsets this advantage. 
This dependence establishes a direct trade-off between memory performance and network infrastructure and can have design implications for future quantum repeater architectures.

Also, classical signaling must be considered an integral part of repeater-network design. Heralding delays determine how long entangled states must remain stored before subsequent operations can be executed, coupling the classical control plane directly to memory decoherence.

The results also indicate that no single ES scheduling
strategy is optimal across all architectural regimes. 
Therefore, future repeater architectures should be able to support the implementation of multiple scheduling strategies rather than
rely on a single fixed one.


Further work may extend the present framework to multiplexed elementary links, where different ES strategies can produce different expected numbers of E2E pairs, introducing an additional trade-off between the quality and the quantity of the distributed entanglement. Other extensions may include the study of heterogeneous chains, platform-specific noise models, and the incorporation of entanglement purification stages. Another interesting direction is to extend the model to fully reactive protocols and determine whether the additional network-state information available to these strategies compensates for the classical signaling delays required to exploit it, under finite-memory conditions.

\bibliographystyle{IEEEtran}
\bibliography{references}


\end{document}